\documentclass{article}

\newcommand{\institute}[1]{\\#1}
\newcommand{\email}[1]{\\ \url{#1}}
\newcommand{\keywords}[1]{
\bigskip
{\bf keywords:} {#1}
\bigskip
}

\usepackage{subfigure}
\usepackage{epsfig}
\usepackage{url}
\usepackage{rotating}

\newcommand{\eg}{{\em e.g.}}
\newcommand{\ie}{{\em i.e.}}

\newcommand{\DP}{$\mathsf{DP}$}
\newcommand{\PC}{$\mathsf{PC}$}
\newcommand{\DPlow}{$\mathsf{DP}\und{low}$}
\newcommand{\DPhigh}{$\mathsf{DP}\und{hi}$}
\newcommand{\DPtwo}{$\mathsf{DP}\und{2}$}
\newcommand{\DPECA}{\DP\ ECA}

\newcommand{\zero}{{\qO ^\LL}}
\newcommand{\one}{{\qX ^\LL}}

\newcommand{\src}{\sr\und{c}}
\newcommand{\srcexp}{{\tilde{\alpha}}\und{c}}

\newcommand{\dsr}{{\Delta_{\sr}}}

\newcommand{\NS}{N\und{s}}

\newcommand{\dinf}{d_\infty}

\newcommand{\DDP}{\delta\und{DP}}

\newcommand{\Tmin}{T\und{min}}
\newcommand{\Tmax}{T\und{max}}

\usepackage{amsmath}
\usepackage{amsfonts}
\usepackage{amssymb}

\newcommand{\Z}{\mathbb{Z}} 			
\newcommand{\N}{\mathbb{N}} 			

\newcommand{\LL}{{\mathcal L}} 			

\newcommand{\fleche}{\rightarrow}		
\newcommand{\function}[3]{ {#1} : {#2} \fleche  {#3} } 	

\newcommand{\set}[1]{\{#1\}}

\newcommand{\und}[1]{_{\mathrm{#1}}}

\newcommand{\TextAnd}{\mbox{ and }}
\newcommand{\TextIf}{\mbox{ if }}
\newcommand{\TextOr}{\mbox{ or }}
\newcommand{\TextOtherwise}{\mbox{ otherwise }}

\newcommand{\ten}[1]{{\ensuremath{10^{#1}}}}
\newcommand{\sn}[2]{{\ensuremath{#1 \times 10^{#2}}}}
\newcommand{\thousand}{\, 000}

\newcommand{\etat}[1]{{\ensuremath{\mathtt{#1}}}}

\newcommand{\qO}{{\etat{0}}}
\newcommand{\qX}{{\etat{1}}}

\newcommand{\OX}{{\qO\qX}}
\newcommand{\XO}{{\qX\qO}}

\newcommand{\qQ}{{\ensuremath{\{\qO,\qX\}}}}

\newcommand{\sr}{\alpha}

\newcommand{\BreakLine}{}

\newcommand{\info}[1]{}

\newcommand{\DPFigure}{
\begin{figure}	
\begin{center}
\includegraphics[width=0.80\textwidth]{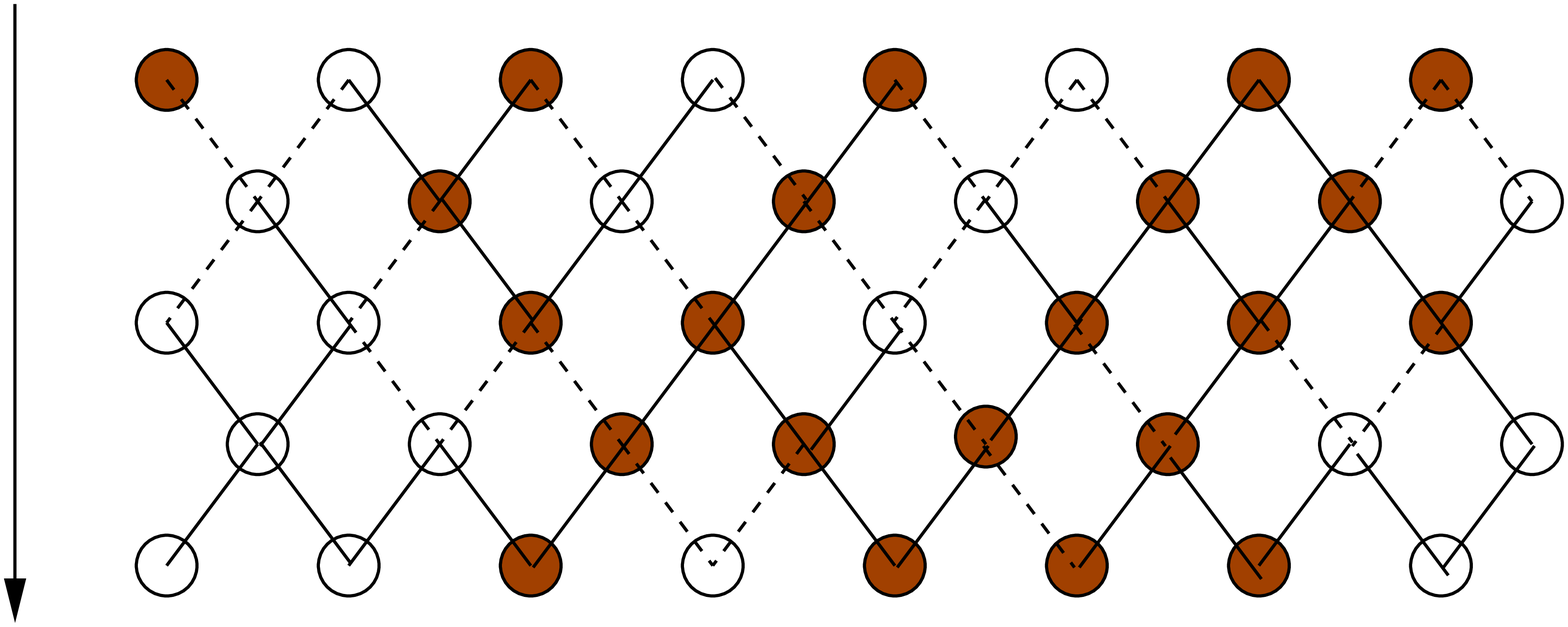}
\end{center}
\caption{Example of directed bond percolation: filled
circles are wet sites, empty circles represent dry sites ; solid
(resp. dashed) represent open (resp. closed) bounds. The links can be
open with a probablity $ p $. The problem is to determine the critical
value $ p_c $ for which the size of the clusters of wet sites diverges.  }
\label{fig:percodir}
\end{figure}
}

\newcommand{\DET}[2]{	
	\includegraphics[width=.24\textwidth, height=!]{Diagrams/DiagECA#1S#2} 
}	

\newcommand{\LineDET}[1]{
\begin{rotate}{90}
$ \longrightarrow $ time 
\hspace{0.6cm}  
ECA #1
\end{rotate}
& \DET{#1}{25} & \DET{#1}{50} & \DET{#1}{75} \\
}

\newcommand{\leg}[1]{$ \sr = #1$}

\newcommand{\FigAlphaDETbis}{
\begin{figure}
\begin{center}
\begin{tabular}{p{0.4cm} c c c }
 & \leg{0.25} & \leg{0.75} &  \leg{0.75} \\
\LineDET{6} 
\LineDET{50} 
\LineDET{178} 
\end{tabular}
\end{center}
\caption{Space time diagrams for ECA 6 (top), ECA 50 (middle) and ECA 178 (bottom).\BreakLine
Synchrony rate is varied : $ \sr = 0.25 $
(left), $ \sr = 0.50 $ (middle),  $ \sr = 0.75 $ (right).\BreakLine 
Time goes from bottom to top; the time factor is rescaled by a factor $ 1/\sr $\BreakLine
(i.e., for  $ \sr = 0.25 $ only time steps that are multiples of 4 are displayed).}
\label{fig:AlphaDET}
\end{figure}
}

\newcommand{\densalpha}[1]{
\includegraphics[width=.46\textwidth]{Figures/DensAlphaECA#1}
}
\newcommand{\kinksalpha}[1]{
\includegraphics[width=.46\textwidth]{Figures/KinksAlphaECA#1}
}
\newcommand{\FigAlphaDensite}{
	
	\begin{figure}[t]
	\begin{center}
	\begin{tabular}{c c }
		\densalpha{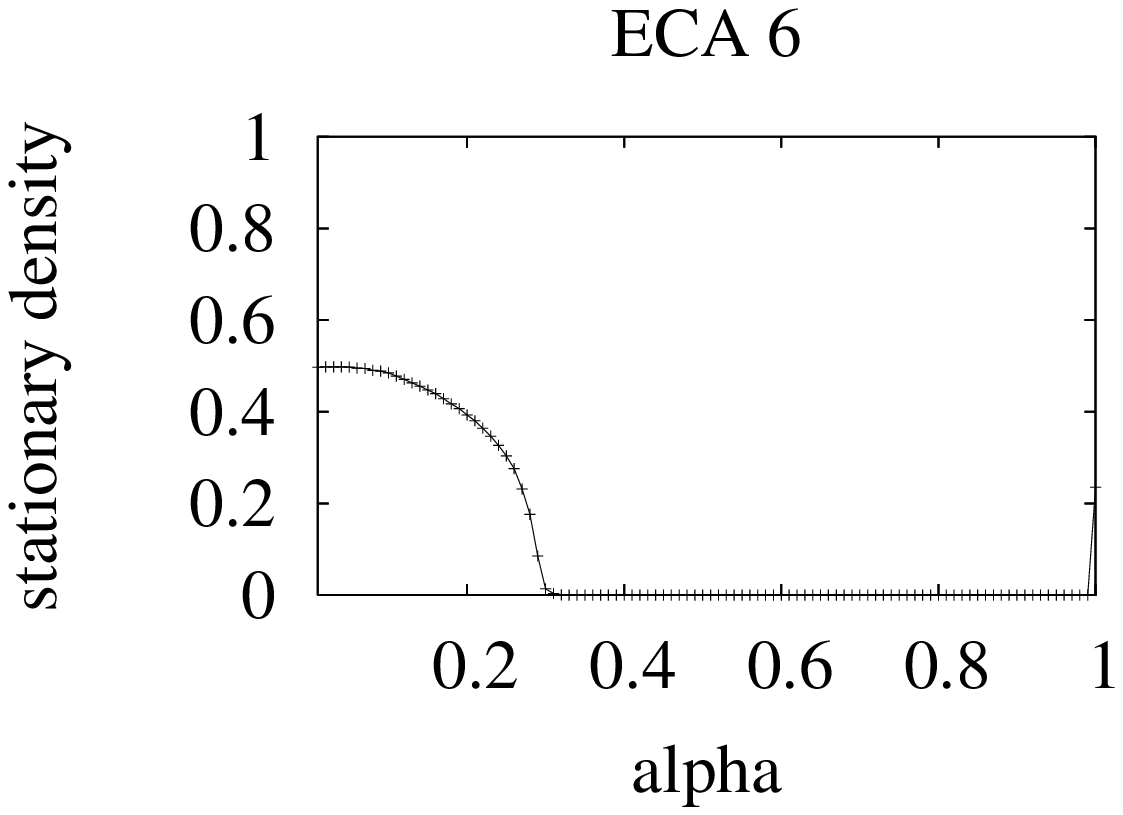}  & \densalpha{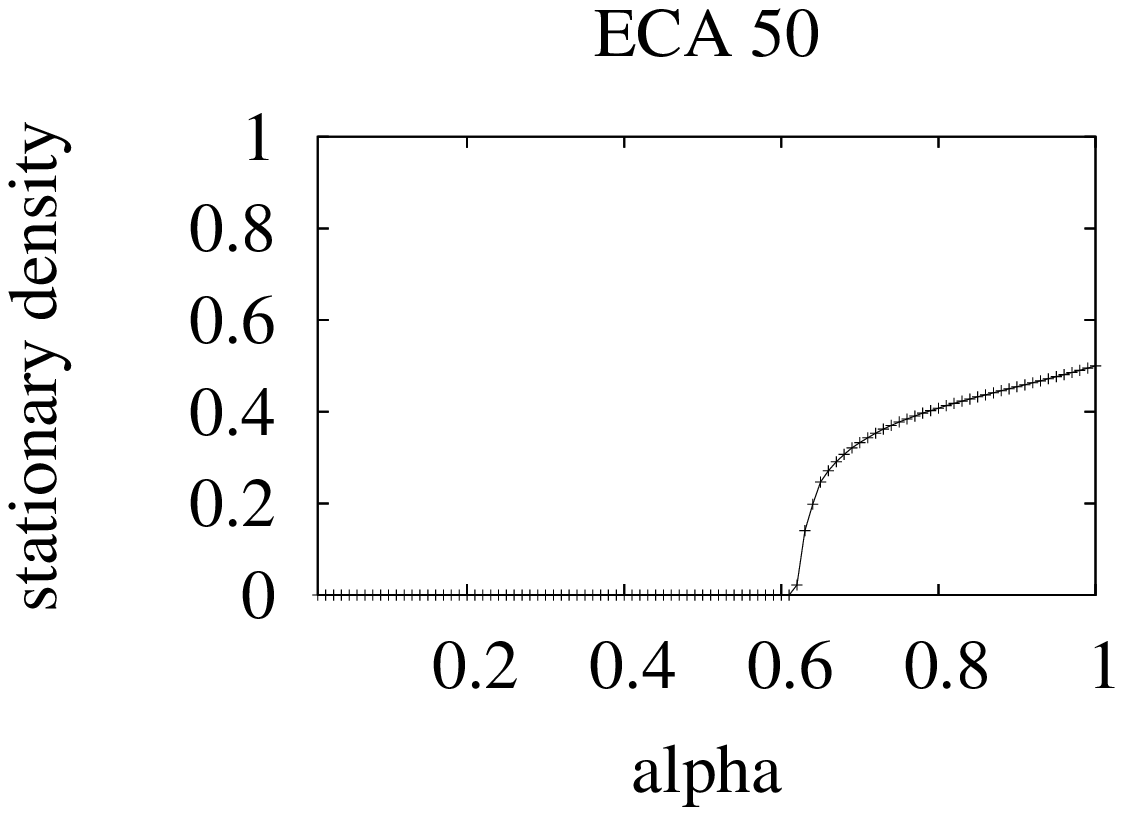} \\
		\densalpha{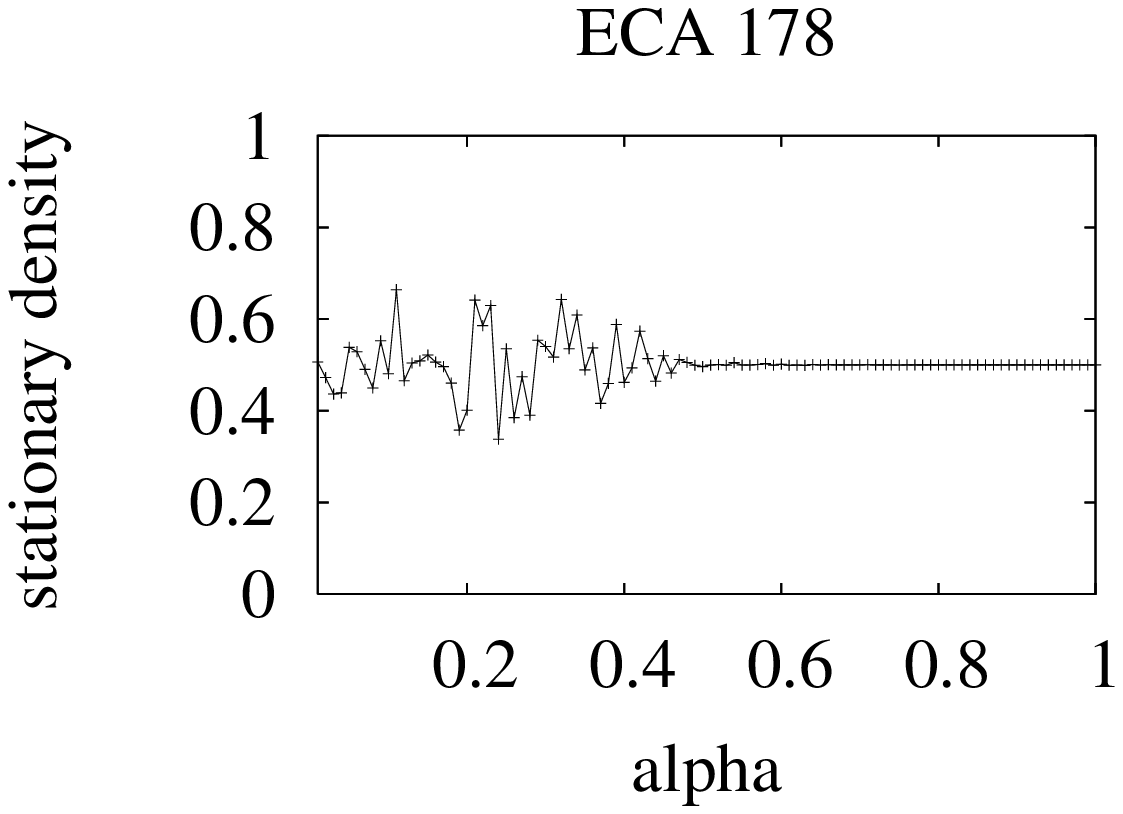} &  \kinksalpha{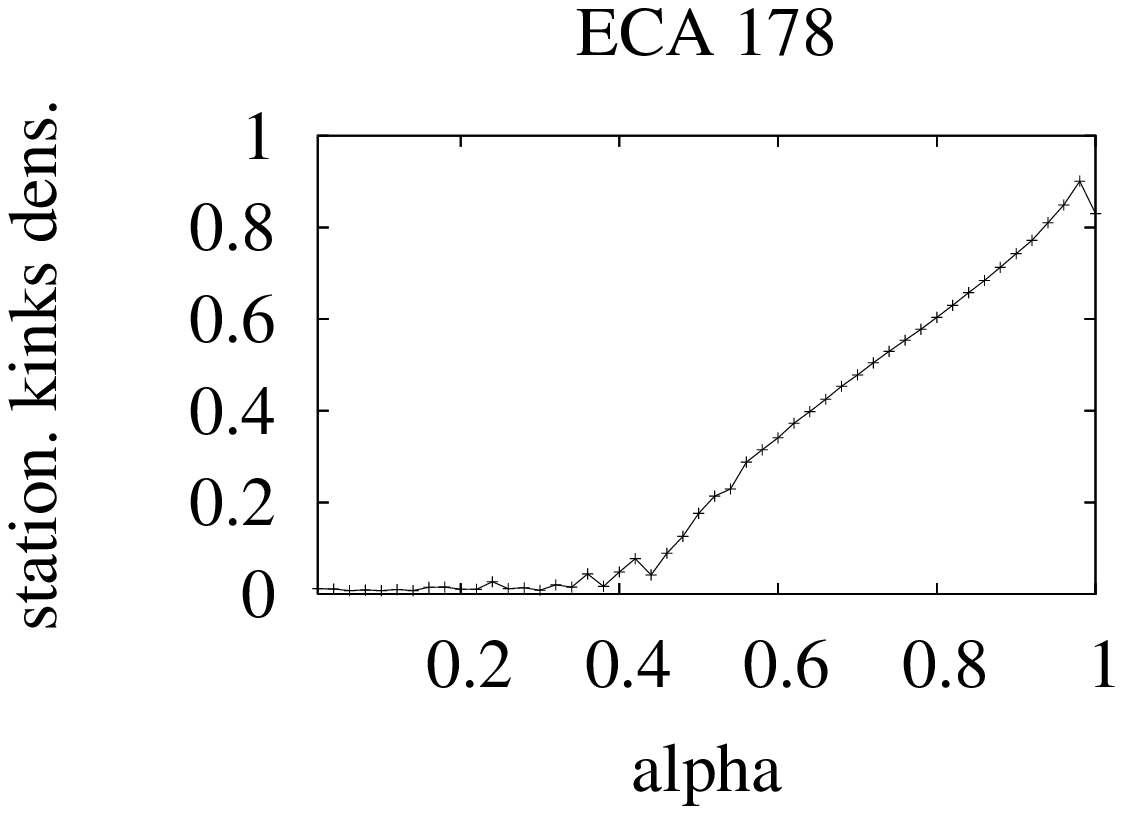} \\
	\end{tabular}
	\end{center}
	\caption{(top) Stationary density versus synchrony
rate for ECA 6 (left) and ECA 50 (right).
(bottom) ECA 178 : stationary density (left) and kinks
density (right) versus synchrony rate.
The discontinuity for ECA 6 and 178 at $ \sr  =1 $ is not an artifact.}	\label{fig:AlphaDensite}

\end{figure}
}

\newcommand{\FigCritical}{
\begin{figure}
	\includegraphics[width=.90\textwidth]{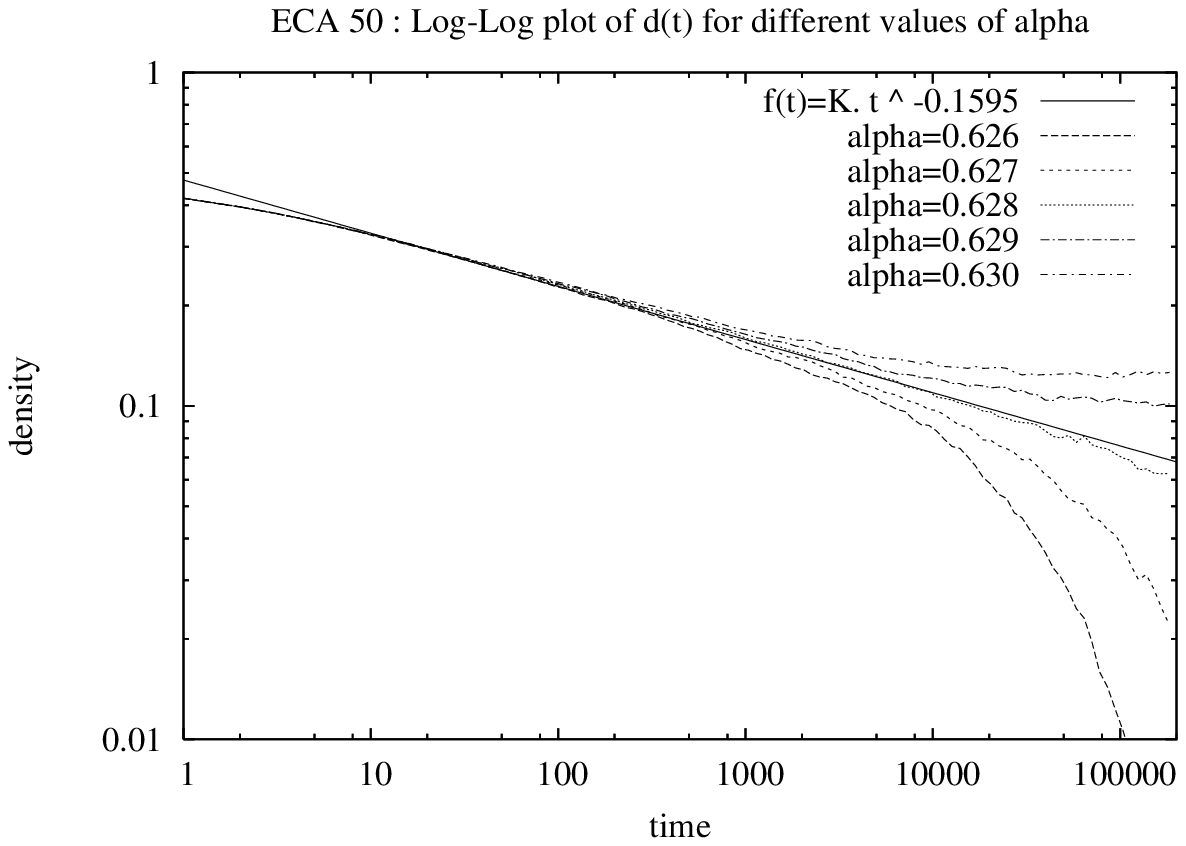}
	\includegraphics[width=.90\textwidth]{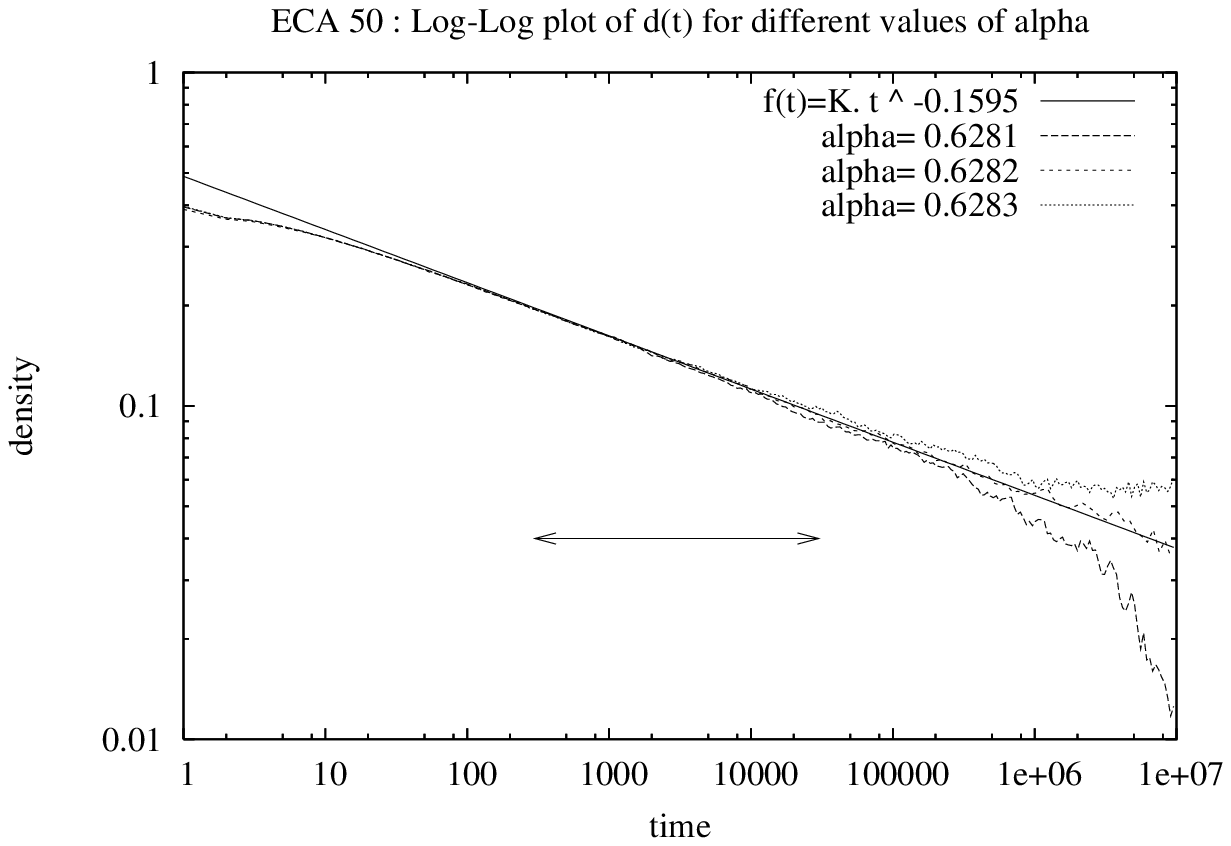}
\caption{
ECA 50: Determination of the critical synchrony rate $\src$.\BreakLine
The straight line has slope $ -\delta\und{DP} =-0.1595 $ and is
plotted for reference.\BreakLine
(above) Averages obtained on $ \NS= 100 $ runs, ring size $ n = 20
\thousand$, sampling time $ T= \sn 2 5 $.
(below)   Averages obtained on $ \NS= 50 $ runs and ring size $ n = 40
\thousand$, sampling time $ T= \ten 7$.
The arrow indicates the interval $ [\Tmin,\Tmax] $ used to perform the
fit to measure $ \delta$.
}
\label{fig:DensityTime}
\end{figure}
}

\newcommand{\FigBetaDensityTime}{
\begin{figure}[p]
	\includegraphics[width=.80\textwidth, height=!]{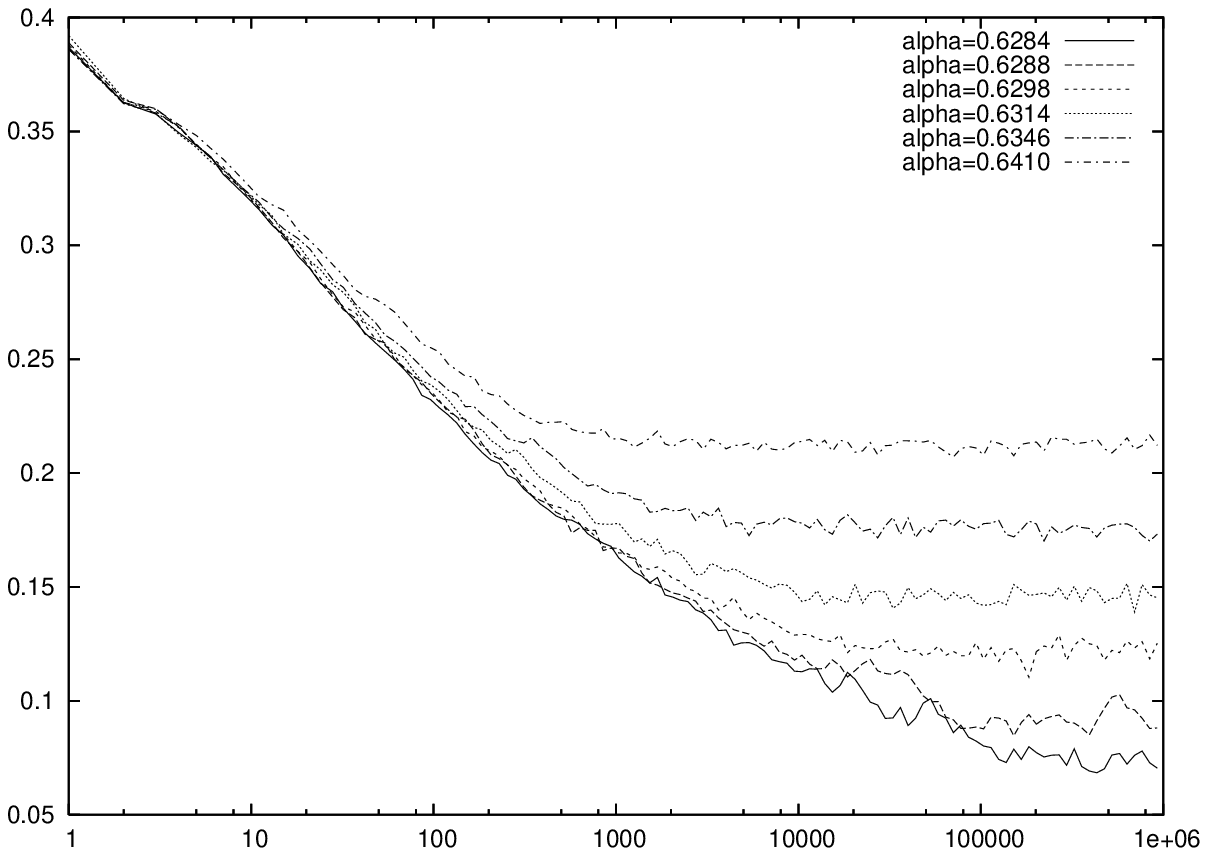}
	\caption{ECA 50: Evolution of the density versus time for $ \sr >
\src = 0.6282$. The ring size is $ n = \sn 2 4 $ ;  averages use $ \NS
= 10 $ runs with a sampling time $ T = \ten 6 $. }
\label{fig:BetaDensityTime}
\end{figure}
}

\newcommand{\BetaLogLog}{
\begin{figure}[p]
	\includegraphics[width=.80\textwidth, height=!]{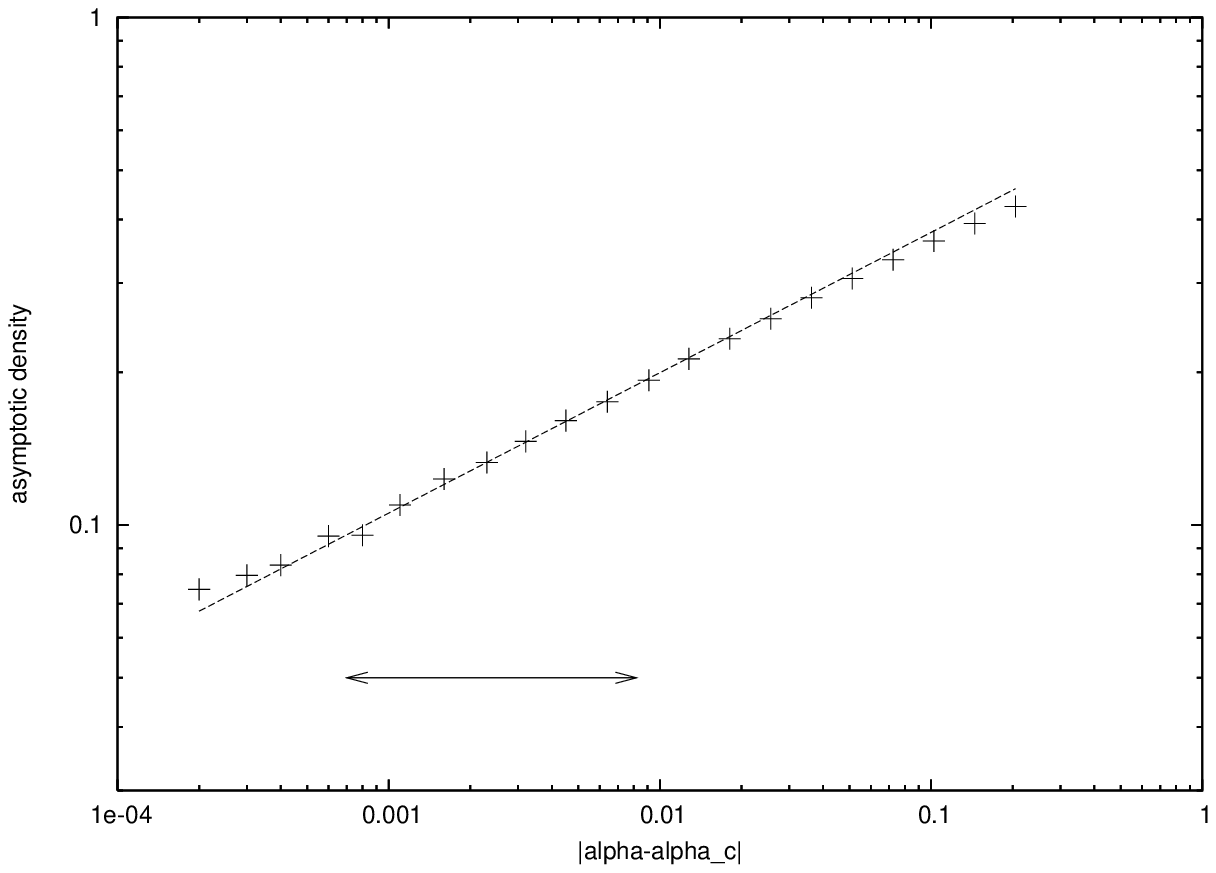}
	\caption{ECA 50: Determination of the critical exponent $
\beta $ using the time decay properties (see text). The
straight line shows theoretical prediction and has slope: $
\beta\und{DP} = 0.2765$. 
Note that both $ x $ and $ y $ axis are displayed in logarithmic scale.
The arrow shows the fit interval used to calculate the $ \beta $
exponent.}
\label{fig:BetaLogLog}
\end{figure}
}

\newcommand{\TimeInt}[4]{\sn{#1}{#2} & \sn{#3}{#4}}

\newcommand{\cntr}[1]{\hfill {#1} \hfill}

\newcommand{\spc}{@{\ \  }}
\newcommand{\result}[7]{#2 & #3 & #4 & #5 &  $ #6\ {({#7})} $ \\}
\newcommand{\TableResultats}{
\begin{table}
\caption{Numerical Results for the \DP\ and \DPtwo\ rules : the digit between parentheses is
uncertainty on the last digit. Compare with $ \delta\und{DP}=0.1595$,
 $ \delta\und{DP2}=0.2856$ and $ \beta\und{DP}=0.276$.}

\centering
\begin{tabular}{| \spc l \spc||\spc l \spc |\spc l  \spc | \spc r  r\spc | \spc c \spc | }
\hline
  \cntr{ECA} & \cntr{$\srcexp$} & \cntr{$  \delta $} & \cntr{$ \Tmin $} & \cntr{$ \Tmax $} & \cntr{$ \beta $}  \\
\hline
\result{BFGH    }{  6 }{ 0.2825  (1) }{  0.160     (3)}{\TimeInt{1}{3}{3}{5}}{  0.27  }{  1     }
\result{BCEFGH  }{ 18 }{ 0.71385 (5) }{  0.157     (3)}{\TimeInt{5}{1}{2}{4}}{  0.27  }{  1     }
\result{BCEGH   }{ 26 }{ 0.47485 (5) }{  0.159     (1)}{\TimeInt{1}{3}{2}{5}}{  0.27  }{  1     }
\result{BDFG    }{ 38 }{ 0.04085 (5) }{  0.160     (4)}{\TimeInt{2}{4}{4}{5}}{  0.27  }{  1     }
\result{BCDEFGH }{ 50 }{ 0.6282  (1) }{  0.158     (3)}{\TimeInt{3}{2}{3}{4}}{  0.27  }{  1     }
\result{BCDEGH  }{ 58 }{ 0.3398  (1) }{  0.161     (4)}{\TimeInt{1}{3}{1}{5}}{  0.28  }{  2     }
\result{BDEH    }{ 106}{ 0.8146  (1) }{  0.157     (2)}{\TimeInt{5}{2}{3}{5}}{  0.28  }{  1     }
\result{BFG     }{ 134}{ 0.0821  (2) }{  0.161     (5)}{\TimeInt{2}{4}{2}{6}}{  0.26  }{  3     } 
\result{BCEFG   }{ 146}{ 0.6751  (1) }{  0.158     (3)}{\TimeInt{5}{1}{1}{5}}{  0.27  }{  2     }
\hline
\result{BCDEFG }{ 178}{ 0.410   (1) }{  0.286    (4)}{\TimeInt{1}{2}{2}{5}}{ ?     }{   ?    }
\hline
\end{tabular}\\

\label{tab:DeltaBeta}
\end{table}
}

\begin{document}

\title{Asynchronism Induces Second Order Phase Transitions in
Elementary Cellular Automata}

\author{Nazim Fat{\`e}s \email{Nazim.Fates@loria.fr}
\institute{
LORIA -- INRIA Nancy Grand-Est \\  Campus Scientifique B.P. 239 \\
54 506 Vandoeuvre-l\`es-Nancy, France.\\
}}

\maketitle

\begin{abstract}

Cellular automata are widely used to model natural or artificial systems. 
Classically they are run with perfect synchrony, {\em i.e.}, the local rule is applied to each cell at each time step. A possible modification of the updating scheme consists in applying the rule with a fixed probability, called the synchrony rate. 
For some particular rules, varying the synchrony rate continuously
produces a qualitative change in the behaviour of the cellular automaton.
We investigate the nature of this change of behaviour using
Monte-Carlo simulations. 
We show that this phenomenon is a second-order phase transition, which
we characterise more specifically as belonging to the directed
percolation or to the parity conservation
universality classes studied in statistical physics.  

\end{abstract}


\smallskip
\noindent {\bf Foreword:} In this article, we only present a limited set of
plots showing our numerical experimentation. The complete set of
graphs can be accessed at:\\
\url{http://www.loria.fr/~fates/Percolation/results.html}\\
The simulations were obtained with the \texttt{FiatLux} CA simulator \cite{FiatLux}.

\pagebreak

\section{Introduction}

With the increase of computing power, cellular automata
(CA) are becoming a popular tool used to simulate various
real-world systems.
While early research was mainly concerned with the study of logical properties of
abstract models \cite{Moore62}, 
many efforts now focus on using CA which closely mimic natural or
artificial phenomena.

Our research aims at studying the
robustness of cellular automata to asynchronous updating, \ie, at evaluating to which extent a small modification of
their updating scheme may perturb their behaviour.
To tackle this problem, we propose to study not only a single model
but a {\em family} of models, where the members are obtained by varying 
the updating scheme \cite{FatesPhD} and keeping the local rule constant.
One simple way of producing such variations is to consider the so
called {\em $\sr$-asynchronous dynamics} \cite{FatesRST06}, 
in which each cell updates its state
with probability $ \sr $, the {\em synchrony rate}, independently at each time step. 

To our knowledge, the problem of comparing the behaviour of
synchronous versus asynchronous CA was first addressed by means of simulation in \cite{Ingerson84},
with a qualitative evaluation of the changes. 
Other experimental studies  followed 
showing that the updating scheme was indeed a key point to study \cite{Ber94,Sch99,Roli02}. 
On the theoretical side, few results have been obtained so far: 
the independence on the order of update history was studied in
\cite{Gacs03,McAu07},
existence of stationary distributions for infinite systems was studied in~\cite{Lou02} and 
a first classification based on the convergence time 
was proposed in~\cite{FatesTCS06,FatesRST06}.

The $\sr$-asynchronous updating was studied experimentally \cite{FatesMorvan05} 
and it was shown that 
the 256 Elementary Cellular Automata (ECA) 
produced various qualitative responses to asynchronism.
We were surprised to observe that, in some cases, a small
variation of the synchrony rate $ \sr $
could trigger a qualitative change of behaviour.  
The present work is devoted to understanding these transitions with an
experimental approach. We extend the
first investigations presented in \cite{FatesDP06}, examining a larger
set of rules with an improved precision. 

In \cite{FatesMorvan05}, we conjectured that the origin of the qualitative change of
behaviour was a phase transition that belonged to the universality class of {\em directed
percolation} (see below). 
We emphasise that this hypothesis was mainly supported by the
visual observation of the space-time diagrams produced near
the critical point. 
The purpose of this article is to investigate this
hypothesis numerically with large Monte-Carlo simulations.

\section{First observations}
\label{sec:First}

This section introduces formal notations and presents a first set of simple experiments.

\subsection{Formal definitions of the model}

Let a ring of $ n $ cells be indexed by $ \LL = \Z/ n \Z $~; a {\em
configuration} is an assignment of state, \qO\ or \qX, to each element of $ \LL $~;
the space of configurations is $ \qQ ^\LL $. 
The {\em density} of a configuration is the
ratio of cells in state~$ 1 $ over the configuration size $ n$. 
The {\em kinks density} is the ratio of the number of $ \OX $ or $ \XO $
patterns over the size of the configuration $ n$.
An {\em Elementary Cellular Automaton} (ECA) is described by a
function $ \function{f}{ \qQ ^3}{\qQ} $ called the {\em
local rule}. ECA are indexed according to Wolfram's usual notation \cite{Wol84}.

We define the {\em $\sr$-asynchronous updating scheme} as the
operation that consists in applying the local rule $ f $ with a
probability $\sr $ or keeping the same state with a probability $ 1 -
\sr $, independently for each cell.
This updating scheme defines a probabilistic global rule which
operates on the random configurations $ (x^t)_{t\in\N} $ 
according to:
$$ 
\forall i \in  \LL,\ 
x_i^{t+1} = 
\begin{cases}
f( x^t_{i-1}, x^t_i, x^t_{i+1} ) &\text{ with probability } \sr \\
x^t_i &\text{ with probability } 1 - \sr 
\end{cases}
$$
By taking $ \sr = 1 $, we fall back on the classical synchronous case
and as $ \sr $ is decreased, the update rule becomes more 
asynchronous while the effect of an update remains unchanged. 
For $ 0 < \sr < 1 $, $ x^t $ is a random configuration that depends on the sequence of cells that
are updated at each time step.

\subsection{Selecting the Rules to Study}

\FigAlphaDensite

\FigAlphaDETbis

In our first work \cite{FatesMorvan05}, we experimentally detected
a first set of rules that showed 
a qualitative change of behaviour when $ \sr $ was varied continuously.  
We re-examined this change of behaviour for the 88
Minimal Representative ECA in more details. 
For each rule, 
we arbitrarily fixed the ring size to $ n = 10 \thousand $ and we
varied $ \sr $ with an increment of $1\%$ from $0.02$ to $1$. 
For each $ \sr$, we started from a
uniform random initial configuration and we measured
the evolution of the density during $ 10\thousand / \sr  $ steps,
putting a limit of $ 100\thousand$ steps for $ \sr <0.1$.
We extracted the average over the second half of the sample, the first
half being used as a transient period to ensure that the system was stabilised.
This average density can be considered as an estimate of the {\em stationary density}, \ie, the limit density that
would be reached if the system size and the transient time grew to infinity.

Figure~\ref{fig:AlphaDensite} shows the estimated stationary
density (or kinks density)
versus~$ \sr$ for ECA 6, 50, and 178 
and Figure~\ref{fig:AlphaDET} shows how the variation of $\sr$
affects the space-time diagrams these rules. 
We observe that for ECA 6 and 50 (respectively ECA 178), a non-zero
density (resp. kinks density) 
corresponds to the existence of stable branching-annihilating
patterns and a zero density (resp. kinks density) 
corresponds to a rapid extinction of these patterns.

Exploring systematically the 88 Minimal Representative ECA, we observed that
ECA 18, 26, 58, 106, and 146 all have plots which are
similar to ECA 50. 
Their behaviour is compatible with a second order
phase transition: there exists a critical value of $ \sr $ such that 
for $ \sr > \src $, the stationary density is non-zero (the {\em active} phase) and for
$ \sr < \src $ the stationary density is zero (the {\em inactive} phase).
The curve is continuous and reaches zero with an infinite slope for the critical value $ \sr_c $.
As our hypothesis is that the nature of the phase transition is {\em
directed percolation} (see Section~\ref{sec:PercoDir}) with an active phase obtained for {\em high} values
of $ \sr$, we call these rules the {\em \DPhigh} ECA.

ECA 6, 38 and 134 displayed a similar behaviour but what is more
surprising is that their phase transition is in an ``inversed''
pattern: the active phase is obtained for {\em small} values of $
\sr$, the inactive phase for large values of $\sr$. 
We call these rules the {\em \DPlow} ECA. 

The case of ECA 178 is also peculiar since the density curve is stable for
large values of $ \sr $ but does not stabilise value for
small values of $ \sr $. This indicates that, 
for this particular rule, the choice of measuring density is not
suitable. 
Instead, if we examine the variation of the {\em
kinks} density, we obtain a smooth curve
with a discontinuity compatible with a second order phase transition.
We call this rule the {\em \DPtwo} ECA. The choice of this name is
justified in the next section.

\section{Phase Transitions}
\label{sec:Context}

In this section, we present a short review of works related
to asynchronous or probabilistic cellular automata and second order
phase transitions.

\subsection{Universality classes}

Many physical or numerical systems exhibit critical phenomena: a
continuous change in the value of a control parameter may produce a
discontinuous response at the macroscopic level.
It is remarkable that near the critical transition, 
the laws governing the evolution of such systems are generally
power laws (see below).
In short, one may understand the origin of these power laws from the
fact that near the transition point, the system has a
self-similar fractal structure and no typical spatial wavelength.
An important property is that the same exponents of the power laws, the {\em
critical exponents}, may be found for different models that do not
necessarily share the same definitions at the microscopic level.

The collection of all models that are described by the same set of
critical exponents is called a {\em universality class}.
Looking at the space-time diagrams produced by the asynchronous ECA, we
originally found similarities with the patterns produced by 
couple map lattices used in hydrodynamics \cite{BerPom94}. These
similarities lead us to identify our phase transition as possibly
belonging to the {\em directed percolation} (\DP) universality class \cite{FatesMorvan05}.

It is beyond the scope of this paper to list all the models that 
belong to the directed percolation universality class  and we refer to
\cite{Hin00, Odo04} for a review.
As far as cellular automata are concerned, directed percolation was observed
in various problems.
To our knowledge, the first CA that was shown to exhibit \DP\ is the
Domany-Kinzel cellular automaton \cite{Kin83, Dom84}. This model is a
tunable probabilistic CA that has the ability to display two
different transitions depending on the tuning of its parameters.
Various other models involving probabilistic CA were
also shown to exhibit \DP\ phenomena (\eg,~\cite{Odor93,Odor96}).

Directed percolation was also identified in problems involving
synchronisation of two instances 
of cellular automata (\eg, ~\cite{Gra99,Rou06}).
To our knowledge, the only example of directed percolation induced by
asynchronism was given by Blok and Bergersen for the famous Game of
Life \cite{Blo99}. The protocol they used to identify the universality
class of the phase transition relied on the measure of a single
critical exponent, the $ \beta $ exponent (see below).

\subsection{Directed Percolation}
\label{sec:PercoDir}

\DPFigure

Percolation problems are studied in the fields of discrete mathematics and statistical physics.
They were initially motivated by the need to model situations in which a fluid
evolves into a porous random medium \cite{Bro57}.
In the classical problem of {\em isotropic percolation}, 
the porous medium is modelled by a regular infinite two-dimensional square lattice.
The nodes of the lattice, or {\em sites}, 
can be either {\em open} (with probability $ q $) 
or {\em closed} (with probability $ 1 - q $) ; 
the links between the sites, or {\em bonds} 
can also be  either {\em open } (with probability $ p $)
or {\em closed} (with probability  $ 1 - p $). 
Starting from an initial set of {\em wet} sites, 
the question is to determine the set of sites that will also be {\em
wet} if the liquid flows into open sites and open bonds. 
The sets of all wet sites connected through closed bonds and sites is called a
{\em cluster}. 

If we set $ q = 1 $ (respectively $ p=1 $), we have the isotropic 
{\em bond} (resp. {\em site}) {\em percolation}.
For the isotropic bond percolation problem, the average cluster size
diverges for the critical value $ p_c = 1/2 $ : we say that the fluids
{\em percolates} through the medium.
{\em Directed} bond percolation is a non-isotropic
variant of the previous model in which links are 
oriented according to a particular direction (see
Figure~\ref{fig:percodir}). 
This models situations in which the fluid can go only in one
direction, for example when submitted to gravity.

We can also formulate directed percolation as a probabilistic dynamical system:
in this case, time plays the role of the non-isotropic dimension.
More formally, if we represent the state of a site $ i \in \N $ at
time $ t $ by  $ s_i^t \in \set{0,1} $ (dry or wet site), 
starting from an initial condition $ s^0 $, 
the states of sites are updated according to the simple rule
\cite{Hin00}:
$$ s_i^{t+1} = 
\begin{cases}
1 & \TextIf [ s_{i-1}^t = 1 \TextAnd {\mathcal L}_i^t(p) = 1 ] \TextOr
[ s_{i+1}^t = 1 \TextAnd {\mathcal R}_i^t(p) = 1 ] \\
0 & \TextOtherwise
\end{cases}
$$
where $ ({\mathcal L}_i^t) $ and $ ({\mathcal R}_i^t) $ are
i.i.d. Bernoulli random variables ; they model the probability
for the left or right bond to cell $ (i,t) $ to be open or closed.

Theory and observations predict that for an {\em infinite}
lattice 
and for a fixed value of $ p $, 
if we start from an initial configuration with all sites in wet state,
the density of wet sites $ d(p,t) $ evolves to a positive
limit for $ p > p_c $ and to a zero limit for $ p \leq p_c $ \cite{Hin00}.
More precisely, if we denote  by $ \dinf( p ) $ the infinite time
limit of $ d(p,t) $, for $ p > p_c $, near the critical point (${p \fleche p_c}$),
the asymptotic density $ \dinf $ diverges from zero by following a power law:
$$
\dinf( p ) \sim ( p - p_c )^{\beta} 
$$
Note that as we have $ \beta < 1 $, the right derivative of $ \dinf( p ) $ has an infinite value at the critical point.
This explains why the transition is experimentally seen as ``abrupt'' 
when $ p $ is varied smoothly.
At the critical point $ p = p_c $, the density vanishes to zero
$ \dinf(p_c) = 0 $ and the decrease follows a power law:\\ 
$$ d( p_c, t) \sim t^{{-\delta}} $$

The values of the two critical exponents $ \delta =  0.1595 $ and $
\beta = 0.2765 $ are known by numerical simulations (the values are given
here with four digits, see \cite{Hin00} for a better precision).
Determining their values analytically is difficult and 
it is so far an open problem to
know whether these exponents are rational numbers.
 There are also other critical exponents that can be used to identify a
universality class but we choose here to focus
only on the measure of $ \beta $ and $ \delta $ exponents 
(following \cite{Mor98,Gra99} for instance).

\subsection{The PC-DP$_2$ class}

In the case where the local rule has two {\em symmetric states},
it is generally observed that the phase transition is either in the {\em $\Z_2$-symmetric directed
percolation} (\DPtwo) class or in the {\em parity-conserving} class (\textsf{PC}). 
Again, we refer to \cite{Hin00,Odo04} for a detailed description of
these two universality classes, their similarities and differences. 
In short, the \PC\ class appears when one considers models
with branching-annihilating random walks with an {\em even} number of
offspring (\eg, \cite{Jen94}). The \DPtwo\ class is observed with models
that introduce two {\em symmetric} states, as it is the case for ECA
178.

It is well known that the two classes coincide for dimension 1
and differ for higher dimensions. An intuitive reason for this
surprising property is indicated by Hinrichsen \cite{Hin00}~: ``active
sites of \PC\ models in $d \geq 2 $ dimensions can be considered as
branching-annihilating {\em walkers}, whereas \DPtwo\ models describe the
dynamics of branching annihilating {\em interfaces} between oppositely
oriented inactive domains''. From the observation of space-time
diagrams of ECA 178 and by analogy with other models, we conjectured 
that the phase transition of this rule belonged to the \DPtwo\
universality class \cite{FatesPhD}.

\subsection{Hypotheses to test}
\label{sec:hypotheses}

The phase transition theory stipulates that there should be two macroscopic parameters, respectively called
{\em the control parameter} and 
{\em the order parameter}, that satisfy the power laws near the
critical point.
For the sake of simplicity, we use the synchrony rate $
\sr $ as a control parameter and the density  $ d $ as an order
parameter. However, note that other possibilities may also be
examined, for example in \cite{Blo99} the authors also use the {\em
activity} (\ie, the ratio of cells in an unstable state) as an order parameter.
For the nine \DPECA\ identified in Section~\ref{sec:First}, we thus expect to measure:

\begin{itemize}
\item $ d(\src,t) \sim  t^{-\delta} $ at the critical point,
\item $ \dinf(\sr) \sim  \dsr^\beta  $ near the critical point, for the active phase,
\end{itemize}
with $ \dsr = \sr - \src $ 
for the \DPhigh\ ECA and $ \dsr = \src - \sr $  for the \DPlow\ ECA.
For the \DPtwo\ rule ECA 178, we take the kinks density as an order
parameter. By contrast with the \DP\ universality class the exponents
of the \PC\ and \DPtwo\ classes are known analytically. In particular
we have $ \delta\und{DP2}= 2 / 7$ for two-dimensional lattices.
(Recall that space and time play a role analogous to the two
dimensions of the grid in the original directed percolation problem.)

\subsection{Protocol and Measures}
\label{sec:Protocol}

The measure of the \DP\ critical exponents is a delicate operation
that generally requires a large amount of computation time. The main difficulty resides in avoiding systematic errors when obtaining statistical data near the transition point. 
Authors have been mislead by their measures
and concluded that a phase transition phenomenon was not in the \DP\
universality class, which has later been proved wrong
by using a different protocol and more precise measures \cite{Jen91,Gra99}.
In order to limit the influence of systematic errors, 
we take the two-step protocol used in~\cite{Gra99}: 
\begin{itemize}
\item We measure the critical synchrony rate $\src $ by varying $ \sr
$ until we reach the best approximation of a power law decay for the
density. This first experiment also allows us to measure the critical exponent $ \delta $. 
\item We measure the stationary density $ \dinf $ as a function of $
| \sr - \src | $ and then fit a power law in order to calculate $ \beta $. 
\end{itemize}
Note that these two steps are not independent since the second operation uses the previously computed value of $ \src $.

The other main concern is to evaluate the influence of finite size effects and metastability.
In the active phase, finite \DP\ systems are in an {\em out-of-equilibrium} state. 
This means that although infinite-size systems have a stable probability
measure for the distribution of states, the finite-size systems used
in the simulations may attain the absorbing state even if they are in
the active phase. There exist many techniques to handle this
problem, for example adding a small amount of noise to prevent the system from
staying indefinitely in the absorbing state. 
In this work we choose to use large
size lattices and verify experimentally that the results are not
influenced by the trajectories that touch the absorbing state (here $ \zero$).

We will present the curves for ECA 50 while only numerical data will
be given for the other \DPECA. Indeed, rule 50 can be written
in the rather simple form:
$$ \forall (a,b,c) \in \qQ^3,\ f(a,b,c) = 
\begin{cases}
\qX - b &\text{ if } (a,b,c) \neq (\qO,\qO,\qO)\\
\qO     &\text{ if } (a,b,c) = (\qO,\qO,\qO)\\
\end{cases}$$
It is possible to see this rule as a one dimensional version of an
``epidemic'' rule: a healthy cell (state \qO) gets infected
(state \qX) if at least one of its neighbour is infected ; once it is
infected, it becomes healthy at the next update of the cell.

\section{Finding the Critical Synchrony Rate $ \src $}

\FigCritical

Following our protocol, we estimate the critical point by 
determining the change of convexity of the density curves in a log-log plot.
The concave function characterises the {\em active} phase as the
asymptotic density evolves towards a non-zero value ; the convex
function corresponds to the {\em inactive} phase as the density
evolves to zero with an exponential decay.

For ECA 50, the previous experiment (Figure~\ref{fig:AlphaDensite}) allowed us to locate the critical
synchrony rate around  $ \sr_c \sim 0.62 $.
Figure~\ref{fig:DensityTime} shows the temporal decay of the density 
curve in a log-log plot, with $ \sr $ varied by increments of $ \ten
{-3} $ around $0.62$.
The change of convexity is found between $ 0.627 $ and $ 0.629 $.
The curves are obtained by averaging the data on $ \NS = 100 $ runs
obtained on a ring size $ n = 20 \thousand $ and a sampling time
of $ T = 200 \thousand $ steps. 

To improve our estimate of $ \sr_c $ to a precision of $ \ten {-4} $,
we extended the sampling time to $ T = \ten 7 $ and 
we increased the ring size to $ n = 40  \thousand$. 
Figure~\ref{fig:DensityTime} shows the two values $ \src^- = 0.6381$  
and $ \src^+ = 0.6383 $ for which we observed the change of convexity.
We repeated the same type of progressive approximations of $ \sr_c$
for all the \DPECA. 
It was always possible to observe the change of convexity when varying
$\sr$ with an increment of $ \ten {-4} $. 
However, the sampling time $ T $, the ring size $ n $ and the number
of samples $ \NS $ had to be adapted to each ECA to ensure a good
stability of the measures (see below). 
The values of the estimated 
critical synchrony rates are reported in Table~\ref{tab:DeltaBeta}.

\subsection{Measurement of $ \delta $}

The value of $ \delta $ is given at the critical point by the slope of the
density curve  in a log-log scale (see Section~\ref{sec:hypotheses}).
We report in Table~\ref{tab:DeltaBeta} the time interval 
$[\Tmin,\Tmax]$ used to compute $\delta$. 
The lower limit $ \Tmin $ is obtained by a rough estimate of the
maximum transient time needed for the system to enter into the power law
regime. 
The upper limit of the interval $ \Tmax $ corresponds to the minimum time for
which the deviation from a power law decay becomes visible 
for the curves $ \src^- $ and $ \src^+ $ that show the change of convexity. 

We bring to the reader's attention the fact  
that it is a difficult problem to estimate the error on $ \delta $. 
Indeed, besides the influence of noise, the value depends on the time
interval $[\Tmin,\Tmax]$  used to perform the fit.
There is to our knowledge no general method for choosing this time interval.
To estimate the error on $\delta$, we computed the different values
obtained when varying $ \sr $ to $ \src^- $ and $ \src^+ $ and by
varying $ \Tmin $ and $ \Tmax $ by a factor $ 2 $.
Numerical values of $ \delta $ are reported in Table~\ref{tab:DeltaBeta} for comparison with $ \DDP $.
Given our estimations of uncertainty, the results show
good agreement with $ \DDP = 0.1595  $.

\subsection{Stability of the measures}

An important question is to determine whether the visual estimations of
the changes of convexity are satisfying and stable.
A simple method to verify this is to estimate numerically the
``linearity'' of the density curves (in a
log-log representation). 
For the sake of simplicity, we used the root mean square error of the
fits as a linearity estimator.
For example, for ECA 50, fitting a power law in the time interval $ [300,
30 \thousand ]$ for $ \src^- = 0.6181 $, $ \src = 0.6282 $, 
$ \src^+ = 0.6283 $ gives an error of 0.16, 0.10, 0.13 (dimensionless units), respectively,
which confirms that the ``most linear'' curve is obtained with $ \src $.

The other point of care concerns the measurement of the critical
exponent~$ \delta $. Recall that we needed to distinguish the
``power law'' part $ [\Tmin,\Tmax] $ of the density curve and the ``departure from power law''
$ [\Tmax, \infty[$, which can be convex or concave.
To which extent does $ \delta$ depend on the choice of $ \Tmin $ and
$ \Tmax $?
Firstly, let us note that the length of transient time interval 
$ [ 0, \Tmin ] $ only depends on the ECA considered while the length of the
``power law'' part of the curve $ [\Tmin,\Tmax] $ is a function of the
distance to the critical point $ |\sr - \src | $: the smaller this
value, the longer the system follows the power law predictions. 
This means that the values of $ \Tmax $ depend heavily on our choice
of increment on $ \sr $. In our experiment, we found
out that it was necessary to take an increment on $ \sr $ as small as
$\ten{-4}$ to ensure that the time interval $ [\Tmin,\Tmax] $ was large enough
to measure $ \delta $ with a good precision.

Secondly, recall that as we use finite size lattices, 
the system is metastable: it eventually reaches the absorbing
state $ \qO ^\LL $, whatever the value of $ \sr $.
In order to infer the infinite-size limit from the simulations obtained
on {\em finite} systems, we need to take $ n $ large enough 
to ensure that the trajectories do not that reach the
absorbing state. 
For all the \DPECA\ but ECA 18, we used $ n = 40 \thousand $ to
satisfy this condition ; for ECA 18, 
it was necessary to take $ n = 80 \thousand $ to prevent the system
from reaching the absorbing state.

Finally, the case of ECA 134 also needs to be underlined as its density
curve has many inflexion points, 
which makes it harder to measure $ \src$ and $ \delta $.

\section{Determination of $ \beta $ }

\FigBetaDensityTime

\BetaLogLog

In this second part of the experiment, we measure the $ \beta $
critical exponent by estimating the
stationary density as a function of $ \dsr = | \sr - \src | $.

\subsection{Measurement of $ \beta $}

There are two different possibilities for varying $ \dsr $: some
authors use linear variation (\eg, \cite{Blo99}) while others use an
exponential variation. As we expect the curve $ \dinf $ versus $ \dsr $
to be linear in a log-log plot, we varied $ \dsr$ with an exponential
increment of $ \sqrt 2 $ from $ \sn{8}{-4} $ to $ 0.512 $, 
which allowed us to obtain equally spaced points in the log-log scale.
Figure~\ref{fig:BetaDensityTime} shows the evolution of the density versus time for different values of $\dsr$.

Following the observations of the previous experiment, we estimated the stationary density $ \dinf(\sr) $ by measuring
$ d(t) $ during $ T = \ten 6 $ steps and by taking the average of $ d $ on the time interval $ [T/2, T]$. 
To limit the influence of the noise, 
we repeated $ \NS = 10 $ times the measure and took the average value.
Note that even though $ T $ can be adjusted as a function of $ \dsr$, 
we prefer to take a fixed value for $ T $ for the sake of simplicity.  
This value was adjusted for the {\em minimal} value of $\dsr$,
which makes it a priori suitable for larger values of $\dsr$ (see Figure~\ref{fig:BetaDensityTime}).

Figure~\ref{fig:BetaLogLog} shows the estimated values of $ \dinf $
as a function of $ \dsr $. The visual comparison with the expected plot
shows a good agreement with the predicted value of \DP. 
For each ECA, we took $ t \in [\sn{20}{-4}, \sn{20}{-3} ] $ as a fit interval. 
The estimated values of $ \beta $ are reported in
Table~\ref{tab:DeltaBeta} ;
as for the $ \delta$ exponent, they show good agreement
with the \DP\ prediction $ \beta_{DP} \sim 0.276 $.

\subsection{Stability of the measures}

To which extent do these measures depend on the setting of
the parameters of the experiment? 
We observe on Figure~\ref{fig:BetaLogLog} that the linear part of the
density curve is limited by two 
competing phenomena. 
Note that the small values of $ \dsr $ are always overestimated.
This can be explained by remarking that, near criticality 
the stationary density vanishes as: $  \dinf(\sr) \sim  \dsr^\beta$.
Note that we also have $ d \sim t^{-\delta}$ near criticality, 
which implies that the time needed to get close to the
stationary density increases exponentially with $ 1 / \dsr $. 
This phenomenon, known as the {\em critical slowing down} (\eg,
\cite{Hin00}),
limits the measurement
of the stationary density for small values of $ \dsr$.

For the higher values of $ \dsr $, the system ``saturates'' and no longer
follows a power law (a density can not be higher than 1).
The deviation from the power law is a phenomenon that is predicted by
theory and that can be studied for its own interest. We
prefer here to restrict our measures to the linear part of the
curve. 
However, some authors advocate that the study of the deviations from
the power law significantly
improves the determination of a universality class \cite{Lub04}.
To estimate the error on $ \beta $, we measured how its value changed
when $ \src $ was varied in the interval $ [\src^-, \src^+ ] $ and
when the fit interval was changed to $ [\sn{30}{-4}, \sn{30}{-3} ] $.
We noticed that the error was mainly due to the uncertainty on $ \src
$ and had an order of magnitude of  $ \ten {-2} $.

\TableResultats

\section{The case of ECA 178}
\label{sec:DP2PC}


For ECA 178, we repeated the protocol described above using the
the kinks density instead of the density as an order parameter.
Our best results were obtained with
$ n =  80 \thousand $, $T= \ten 8 $ and $ \NS = 50 $:
we located a change of convexity between 
$ \src^- = 0.409 $ and  $ \src^+ = 0.411 $.
However, by contrast with the previous \DPECA, this change of convexity
was much more difficult to observe. 
This well-known difference is explained by the algebraic decay of the
order parameter in \PC\ or \DPtwo\ models.
This produces straight lines
in a log-log scale  which makes changes of convexity less visible 
than the exponential decay observed for the \DPECA.
This explains why it is difficult to estimate $ \src $ with a better
precision despite using a long computation time. 

In the time interval $ [100, 20 \thousand] $, the fit for the curve $ \src =
0.410 $ gives a slope of $ \delta = 0.2860 $, which is close to the
predicted value $ \delta\und{DP2} \sim 0.2856 $.
To estimate the error on $\delta$, we repeated the fit with the
same time interval, 
but with the curves obtained with  $ \src^- $ and $ \src^+ $. 
We found that $ \delta $ varied less than $ \sn 4 {-3} $.
As the precision on $ \delta $ is good 
while the precision on $ \src $ is relatively poor for this rule, we did not measure the critical exponent $ \beta $. 
A possible way of improving our measure of the critical point $ \src $
consists in using a different experimental protocol, for example
measuring the ``dynamical''
exponents obtained by starting from an initial condition close to the
absorbing state.

\section{Discussion}
\label{sec:Discussion}

We experimentally investigated how qualitative 
changes of behaviour were triggered by
gradual variations of the synchrony rate in asynchronous cellular
automata. 
The results show good evidence that these phenomena are second
order phase transitions which belong to the directed percolation
(\DP) and parity conservation (\PC- \DPtwo) universality classes.  

From a practical point of view, the main limitation for identifying
the phase transitions was the
huge amount of computation time required for measuring the critical exponents
: for each ECA, we used approximately $ \ten{15} $ computations of the
local rule in order to determine these exponents with two
or three digits. This task represented several months of calculus
with several personal computers. It calls for further studies on how
to improve the computation process (\eg, the generation of random
numbers) or on how to take advantage of the use of massively parallel computing
devices.

\newcommand{\itm}[1]{{\bf(#1)}}
\subsection{The origin of phase transitions}

The observation of the synchronous behaviour of the rules we studied
indicate that there is certainly no straightforward relation with the
existing classifications. For example, ECA 50 is ``periodic'' (or
Wolfram class II) while ECA 18 is ``chaotic'' (or Wolfram class III). 
This indicates that some particular
asynchronous cellular models obey the {\em same} macroscopic laws near their
transition points despite of having different definitions at the
microscopic cell-scale.
In other words, the asynchronous updating of cellular systems unveils
another type of complexity which still needs to
be understood.

A first path for discussing these results is to consider the famous 
conjecture by Janssen and Grassberger 
(\eg, see \cite{Hin00} for a short presentation). It states that a
model should belong to the \DP\ universality class {\em if} it satisfies the following criteria:
\itm{a}~the model displays a continuous phase transition from a
fluctuating active phase to a {\em unique} absorbing state, 
\itm{b}~it is possible to characterise the phase transition by a
positive one-component order parameter, 
\itm{c}~the dynamics of the model is defined by short-range process, 
and \itm{d}~there exists no additional symmetries or quenched
randomness (\ie, small and stable topological modifications).

For all the \DPECA, we saw that condition \itm{a} was fulfilled  with $ \zero $ as the
absorbing state. 
It is interesting to note that configuration $ \zero $ is a fixed point for all the \DP\ rules,
but ECA 134 and 146 also have $ \one $ as a fixed point. 
This confirms that the
informal notion of ``absorbing state'' can not be trivially identified with the
fixed point mathematical property. 
Condition \itm{b} was fulfilled by the density or by the kinks density. 
Condition \itm{c} is true by definition of cellular automata.

Analysing condition \itm{d} is more interesting since we see a difference between
{\em space} symmetry, which is possessed by ECA 50 for example, and {\em state}
symmetry (\ie, invariance under \qO\ and \qX\ exchanging), which is
absent for all the \DPECA. 
This may also explain why ECA 178, which has both symmetries, has
a peculiar behaviour and was found in the \PC-\DPtwo\ universality class.
Note that conditions \itm{b}, \itm {c}, \itm{d} can be easily verified but
a challenging question consists in explaining why only
a small fraction of the 256 ECA also verify condition \itm{a} and thus exhibit \DP\ behaviour.

\subsection{Perspectives}

At this point, it is still an intriguing question to
understand the origin of out-of-equilibrium phase transitions in cellular automata, and more generally, in particle systems. 
The models we exhibited count among the simplest models that display
out-of-equilibrium phase transitions.
In that respect, they can be used as a good basis for exploring 
the origin of phase transitions by analytical means or by numerical
simulations.

A further step for understanding asynchronous cellular automata is
to make a ``reduction'' between the \DP\ rules, \ie, to show that if one of
them exhibits such a phase transition, then the others behave similarly.
Another step is to extend such a reduction to other well-studied
systems such 
as the Domany-Kinzel probabilistic CA~\cite{Dom84}. 
This implies uniting the two types of
phase transitions we observed: recall that going from the inactive to active phase was obtained either by {\em increasing} (\DPhigh\ ECA) 
or by {\em decreasing} (\DPlow\ ECA) the control parameter.
To our knowledge, it is the first example where these two-way
transitions are observed.

A more ambitious possibility for unifying the study of all the \DPECA\ is to
consider the larger set of {\em probabilistic} elementary cellular automata, which
include the $\sr$-asynchronous ECA. 
In this space, which is homeomorphic to $ [0,1]^8 $, 
the problem is to determine whether the phase
transitions can be explained in terms of crossing of a hypersurface
(see also~\cite{Chate88}).

Our view is that there exists a wide range of problems that could
benefit from the study of cellular systems with out-of-equilibrium phase transitions.
For example, in biology, can we explain the trigger of the self-organisation phase in cellular
societies by using similar models \cite{FatesMorvan05, Ber03}?
In an engineering context, 
can we use phase transitions to design systems that change their
behaviour without any external control ?

{
\bibliographystyle{plain}
\bibliography{CAbiblio,DirectedPercolation,Fates}
}

\section{Acknowledgements}

\noindent {The author expresses his acknowledgements to anonymous
referees as well as H. Berry, A. Boumaza, O. Buffet, A. Dutech,
E. Flach, N. Paul and
B. Scherrer for their careful reading of the manuscript.}

\end{document}